\documentstyle[a4wide,epsf,11pt,titlepage]{article}

\newcounter{nref}
\newcommand{\bbib}{%
  \renewcommand{\refname}{\large\bf References}%
  \begin{thebibliography}{99}%
  \setcounter{enumiv}{\thenref}}
\newcommand{\ebib}{%
  \end{thebibliography}%
  \setcounter{nref}{\arabic{enumiv}}}
\newcommand{\head}[3]{%
  \setcounter{nref}{0}%
  \thispagestyle{empty}%
  \section*{\LARGE\bf #1}%
  \stepcounter{section}%
  \addcontentsline{toc}{section}{#1}%
  \large\itshape%`````
  #2\\\vspace{0.1pt}\\%
  #3%
  \normalsize\upshape%
  \bigskip}

\begin{document}

%----------------------------------------------------------------------

\head{Investigating the Flame Microstructure \\in Type Ia Supernovae}
     {F.K.\ R\"opke, W.\ Hillebrandt, J.C. Niemeyer}
     {Max-Planck-Institut f\"ur Astrophysik,
     Karl-Schwarzschild-Str.~1, D-85740 Garching, Germany}

\subsection*{Introduction}
Type Ia Supernovae (SN Ia in the following) are the most accurate 
cosmological distance indicators at the moment.  Through light curve
shape corrections accuracies of better than 10\% in distance are
claimed. In addition, claims of an accelerated cosmic expansion and
thus a new constituent of the universe are mainly based on supernova
observations. It is, however, evident that an understanding of the
explosion mechanism is needed in order to validate such
interpretations of the data and to control possible evolutionary effects.

One way to perform such investigations is by means of numerical
simulations. These simulations should be self-consistent,
independent of any phenomenological parameters, and based on models
that are in accord with observational constraints. However,
the main problem with the numerical approach is the vast range 
of relevant length scales, 11 orders of magnitude from the flame 
width to the radius of the star, that remains unresolvable in
multidimensional simulations. Following \cite{fritz.1},
attempts to overcome this difficulty can be classified into large scale
calculations (LSCs) and small scale calculations (SSCs). The former try
to simulate SN Ia on scales of the radius of the exploding white dwarf
star relying on assumptions on the microscale physics, whereas the
latter are used to study the flame dynamics through
a small window in scale space. In this manner it is possible to
isolate specific physical effects and eventually to include the
information gained here into LSCs. Examples of LSCs are given in
M.~Reinecke's contribution to these proceedings. The
present work is concerned with SSCs to investigate the dynamics of
thermonuclear flames in SN Ia explosions.

\subsection*{The astrophysical model}
The SN Ia model adopted in the present study describes them 
as thermonuclear explosions of  Chandrasekhar-mass 
C+O white dwarf stars. In the standard model nuclear burning is
ignited near or at the star's center. At the typical temperatures 
of around $10^{10}\;$K the thermonuclear burning rate scales with 
the temperature as $\dot{S} \propto T^{12}$. Hence the burning 
takes place in a very thin region which is called a flame.

Flames can travel in two distinct modes: as a supersonic compression
(shock) wave in the detonation mode and as a subsonic conductive wave in
the deflagration mode. A pure detonation model for SN Ia can be
ruled out because the outcome does not match the observed spectra. A
more promising model (see M.~Reinecke's contribution) is one in which the
explosion starts out as a (slow) deflagration, is accelerated by
turbulence, and possibly undergoes a transition to a detonation. The 
questions arising here concern the mechanism of the generation of
turbulence as well as the likelihood of a deflagration-to-detonation
transition. They are still not answered and are the motivation to study 
thermonuclear flames at microscopic scales.

In this work we will investigate length scales much larger than the 
diffusive flame width (less than one millimeter for C+O white dwarfs 
\cite{fritz.2}) but much smaller than the stellar radius.
On such scales the so-called discontinuity approximation holds which
describes the flame as a discontinuity in temperature and density. This
picture does not resolve the inner structure of the flame and it is
therefore necessary to prescribe the laminar burning velocity as 
taken from direct numerical simulations, e.g. \cite{fritz.2}.

\subsection*{Some theoretical considerations}
A laminar flame is subject to various instabilities. In the context of
SN Ia explosions the most important ones are the Rayleigh-Taylor
instability, the Kelvin-Helmholtz instability, and the Landau-Darrieus
instability. While the first two are more important on large scales,
the Landau-Darrieus (LD) instability acts unconditionally on all
scales provided the flame can be described by the discontinuity
approximation. This instability shall be investigated here, and its
cause will be described briefly. 

As was first discussed by Landau \cite{fritz.3} and Darrieus 
\cite{fritz.4} the cause of this instability is 
the refraction of the streamlines of the flow at the density jump of
the flame. Consider a flame front that is perturbed from its originally
planar shape. In the vicinity of a bulge of the perturbation the flow
tubes are broadened. This leads to a decrease of the local fluid
velocity with respect to the velocity far away from the
front. Therefore the burning velocity $s_l$ of the flame is higher than the
corresponding local fluid velocity leading to an accrual of the
bulge. The opposite holds for recesses of the perturbed front. In this
way the perturbation keeps growing. By means of a linear stability
analysis Landau found for the growth rate of the perturbation
amplitude
\begin{equation}
\label{fritz.eq1}
  \omega_{\mathrm{LD}} = k s_l \frac{\mu}{1 + \mu}
  \left(\sqrt{1 + \mu - \frac{1}{\mu}} - 1\right),
\end{equation}
where $k$ denotes the perturbation wavenumber and $\mu =
\rho_u/\rho_b$ is the expansion across the flame front. In white dwarf
matter at densities relevant for SNe Ia, the existence of the LD
instability was first demonstrated numerically in \cite{fritz.4b}. 

\begin{figure}[ht]
  \centerline{\epsfxsize=0.5\textwidth\epsffile{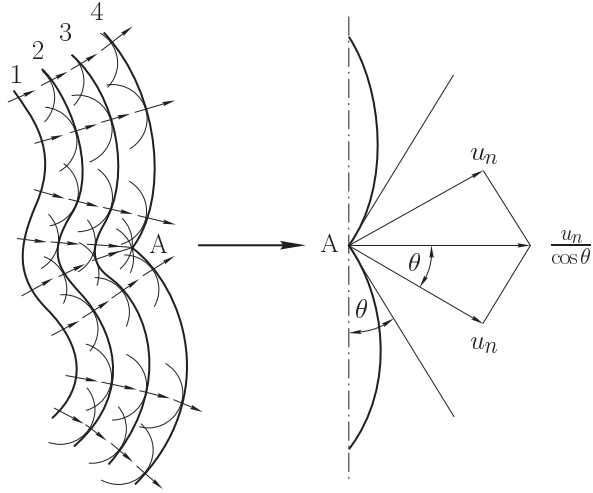}}
  \caption{Nonlinear effect stabilizing the flame front (adopted from
    \cite{fritz.5}).}
  \label{fritz.fig1}
\end{figure}

However, the LD instability does not lead to unlimited growth 
of the amplitudes but is limited in the nonlinear regime.
Fig.~\ref{fritz.fig1} illustrates this effect which can
be described in a geometrical way (following \cite{fritz.6}): The
propagation of an initially sinusoidally perturbed flame is first 
determined by Huygens' principle. But once a cusp forms at point A  
Huygens' principle breaks down and the flame propagation
enters the nonlinear regime. The propagation velocity at the cusps
exceeds $s_l$. This leads to a stabilization of the flame in form of 
cells which propagate with a velocity not much higher than  $s_l$. 
Therefore the LD instability is usually ignored in SN Ia simulations. 
An analytical investigation of the stabilizing effect is given in 
\cite{fritz.7} and is applied to the context of SN Ia in \cite{fritz.8}.
But there exist the possibility that nonlinear stabilization is destroyed
under certain conditions (e.g. certain densities; interaction with
turbulent velocity fields). In the following section we present
a numerical method that will allow us to study this question.

\subsection*{Numerical methods}
The fluid dynamics is described by the reactive Euler equations that
are discretized on an equidistant Cartesian grid. To solve these
equations a Godunov scheme is applied using the piecewise parabolic
method (PPM) \cite{fritz.9}. For this we employ the {\sc Prometheus}
implementation \cite{fritz.10} with directional splitting in two
dimensions.

The equation of state is that of white dwarf matter, including an
arbitrarily degenerate and relativistic electron gas, local black body
radiation, an ideal gas of nuclei and electron-positron pair
creation/annihilation. 

Thermonuclear burning as considered here takes place at fuel densities
above $10^7\,\mathrm{g}\,\mathrm{cm}^{-1}$ and terminates into nuclear
statistical equilibrium, consisting mainly of $^{56}$Ni. Because of the
high computational costs of a full reaction network and because the
objective of our SSC is to study flame dynamics rather than to provide
an exact nucleosynthetic description we simplified the burning to an
net reaction, $14\;^{12}$C$\longrightarrow 3\;^{56}$Ni, which gives an
energy release of $7\cdot 10^{17}\,\mathrm{erg}\,\mathrm{g}^{-1}$.

\begin{figure}[ht]
  \centerline{\epsfxsize=0.8\textwidth\epsffile{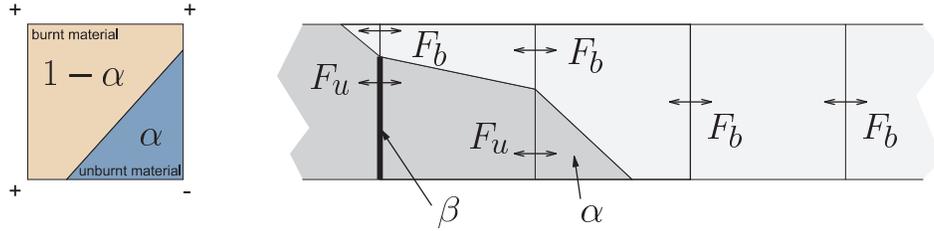}}
  \caption{Determination of the unburnt volume fraction $\alpha$ (l.h.s.);
           flux-splitting method, adapted from \cite{fritz.11} (r.h.s.)}
  \label{fritz.fig2}
\end{figure}

The flame is described using the {\em level set technique},
associating the flame front with the zero level set of a signed
distance function $G$. Details of the implementation of the level set
method can be found in \cite{fritz.11}. Without additional measures,
however, the flame would smear out over 3-4 cells of the 
computational grid. This is not acceptable for our
purposes and therefore we apply the {\em in-cell
reconstruction/flux-splitting scheme} developed by Smiljanovski et
al. \cite{fritz.12}. This algorithm makes use of the fact that
geometrical information on cells containing burnt and unburnt material
(``mixed cells'') can be obtained from the intersection of the zero
level set of the $G$-function (representing the flame) with the
interfaces of the mixed computational cells. In this way one can
designate the volume fraction $\alpha$ of the unburnt cell part
(Fig.~\ref{fritz.fig2}, l.h.s.).
The conserved quantities $U$ are now interpreted
as a linear combination of unburnt and burnt parts of each cell
$$
  \overline{U} = \alpha U_u + (1 - \alpha) U_b.
$$
Together with the Rankine-Hugoniot jump conditions and the Rayleigh
criterion for the flame front, the equation of state, the jump
condition for the velocity component normal to the front, and the
prescribed burning velocity one obtains a system of equations which
allows to compute state variables of unburnt and burnt 
matter in each mixed cell from the mean values. To describe the energy 
generation and species conversion consistently it is necessary 
to treat the source terms in this reconstruction implicitly. 
The reconstruction equation system is solved with Broyden's method.

The knowledge of the values of the burnt and unburnt quantities makes
it possible to treat the hydrodynamic fluxes as linear combinations
of ``unburnt'' and ``burnt'' fluxes weighted with corresponding parts
$\beta$ of the cell interfaces (see Fig.~\ref{fritz.fig2}, r.h.s.). These
partial fluxes are calculated separately. Therefore it is guaranteed
that for instance  no unburnt material can flow through the third cell
interface from the left in Fig.~\ref{fritz.fig2}, r.h.s. This feature
prevents the flame from smearing out and enables us to describe it as a
sharp discontinuity. 

\subsection*{Some preliminary results}
In order to study the LD instability the flame evolution has been simulated
for two values of the fuel density, namely for $\rho_u = 5\cdot
10^7\,$g$\,$cm$^{-1}$ and for $\rho_u = 5 \cdot 10^{8}\,$g$\,$cm$^{-1}$. The
corresponding expansion coefficients across the front are $\mu =
2.410$ and $\mu = 1.619$, respectively. The simulation runs were
performed on a grid of 100 $\times$ 100 cells
with a cell length of 500$\,$cm. The physical domain was periodic in
the $y$-direction. On the left boundary of the domain an outflow 
condition was enforced and on the right boundary we imposed an inflow 
condition with the unburnt material entering the grid with the laminar
burning velocity $s_l$.

\begin{figure}[ht]
  \centerline{\epsfxsize=0.95\textwidth\epsffile{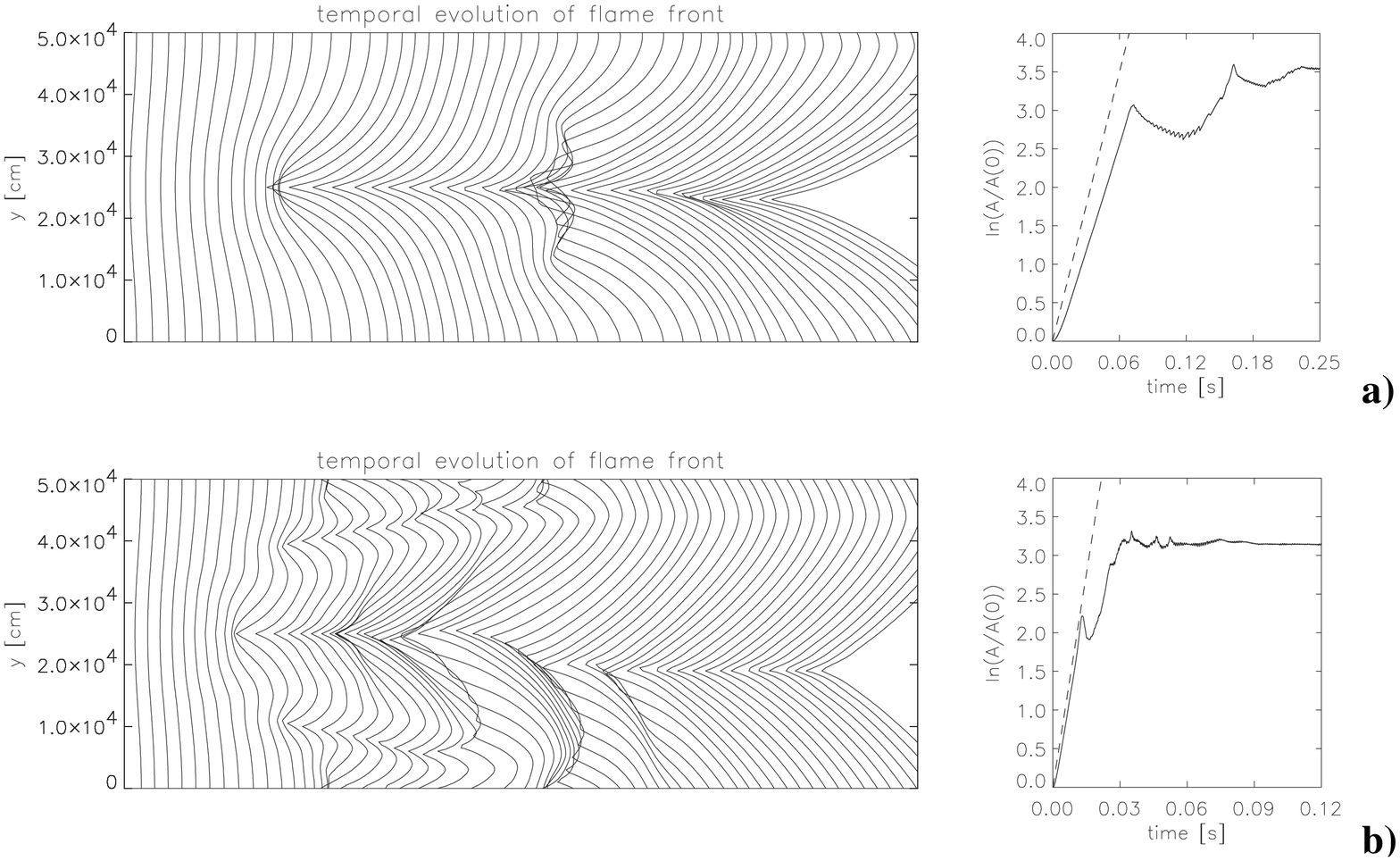}}
  \caption{Flame front evolution. a) $\rho_u=5 \cdot 10^7
  \mathrm{g}\,\mathrm{cm}^{-3}$, each contour corresponds to a time
  step of $\Delta t = 2.46\cdot 10^{-3}\,$s. b) $\rho_u=5 \cdot 10^8
  \mathrm{g}\,\mathrm{cm}^{-3}$, $\Delta t = 1.23\cdot
  10^{-3}\,$s. The plots on the r.h.s. show the
  growth of the perturbation amplitude over time (solid) and the
  prediction of Landau's linear stability analysis (dashed). }
  \label{fritz.fig3}
\end{figure}

Fig.~\ref{fritz.fig3} shows the temporal evolution of the flame
fronts. The flame was initially perturbed in a sinusoidal way with a
wavelength of $\lambda_{\mathrm{pert}} = 5 \cdot 10^4\,$cm and an
amplitude of 200$\,$cm (0.4 grid cells). Each contour is shifted
artificially in $x$-direction for better visibility. As
expected, in the beginning the perturbation grows exponentially, 
and the growth rate of the perturbation amplitude meets the 
theoretical prediction (\ref{fritz.eq1}). As the flame
evolution enters the nonlinear regime the formation of a cusp sets
in. The front develops a cellular
structure. At the troughs the cells split forming new cusps. The new
cells move downwards the initial cusp and disappear. Finally this
merging process results in a single domain-filling cusp/trough
structure which then propagates steadily forward.

We do not reproduce the result of \cite{fritz.4b} that nonlinear
stabilization fails at  $\rho_u = 5\cdot 10^7\,$g$\,$cm$^{-1}$. A
detailed discussion of this discrepancy will be presented elsewhere.

\subsection*{Summary and outlook}
The presented model was proven to be applicable for studies of the
microscale behavior of thermonuclear flames in the discontinuity
approximation. The response of the flame to perturbations of its
planar shape meets the theoretical expectations. Early stages of the
flame evolution are consistent with Landau's stability analysis. The
later evolution of the flame is nonlinear, it stabilizes in a cellular
structure. Whether the final outcome is always a single domain-filling
cell has to be studied in larger simulations. With this new code we 
are now in a position to investigate the interaction of thermonuclear 
flames with imprinted velocity fields such as turbulence and to study 
the question of nonlinear stability. Moreover, if possible, we want
to apply the in-cell reconstruction/flux-splitting technique to
full two-dimensional LSCs of SN Ia explosions. 

\subsection*{Acknowledgements}
We would like to thank Martin~Reinecke, Heiko~Schmidt, and
Sergei~Blinnikov for valuable discussions.

%----------------------------------------------------------------------


\bbib
\bibitem{fritz.1} J.C.~Niemeyer and W.~Hillebrandt, in
  \textit{``Thermonuclear Supernovae''}\/, eds. Ruiz-Lapuente et al.,
  Kluwer Academic Publishers, Dordrecht 1997, p. 441.
\bibitem{fritz.2} F.X.~Timmes and S.E.~Woosley, ApJ {\bf 396} (1992)
  649.
\bibitem{fritz.3} L.D.~Landau, Acta Physicochim. URSS {\bf 19} (1944)
  77.
\bibitem{fritz.4} G.~Darrieus, communication presented at {\it La
    Technique Moderne} (1938), unpublished.
\bibitem{fritz.4b} J.C.~Niemeyer and W.~Hillebrandt, ApJ {\bf 452} (1995)
  779.
\bibitem{fritz.5} Ya.B.~Zeldovich, G.I.~Barenblatt, V.B.~Librovich,
  and G.M.~Makhviladze, {\it The Mathematical Theory of Combustion and
    Explosions}, Nauka, Moscow (1980)
\bibitem{fritz.6} Ya.B.~Zel'dovich, Journal of Appl. Mech. and
  Tech. Physics {\bf 1} (1966) 102. 
\bibitem{fritz.7} G.I.~Sivashinsky, Acta Astronautica {\bf 4} (1977)
  1177.
\bibitem{fritz.8} S.I.~Blinnikov and P.V.~Sasorov, Phys. Rev. E {\bf 53} (1996)
  4827.
\bibitem{fritz.9} P.~Colella and P.R.~Woodward, J. Comput. Phys. {\bf 59} (1984)
  174.
\bibitem{fritz.10} B.A.~Fryxell, E.~M\"uller, and W.D.~Arnett, MPA
  Preprint {\bf 449} (1989) 
\bibitem{fritz.11} M.A.~Reinecke, Ph.D. thesis, Technische
  Universit\"at M\"unchen (2001)
\bibitem{fritz.12} V.~Smiljanovski, V.~Moser and R.~Klein,
  Comb. Theory and Modeling {\bf 1} (1997)
  183.
\ebib
\end{document}